\documentclass[apj]{emulateapj}

\newcommand{\kev}{keV}
\newcommand{\etal}{et al.}
\newcommand{\chandra}{\textit{Chandra}}
\newcommand{\xmm}{\textit{XMM-Newton}}
\newcommand{\spitzer}{\textit{Spitzer}}

\slugcomment{Accepted to the ApJ}

\shorttitle{AGN and the Cosmic Background Radiation}
\shortauthors{Ballantyne \& Papovich}

\usepackage{times}
\usepackage{graphicx}

\begin{document}

\title{On the Contribution of AGN to the Cosmic Background Radiation}


\author{D. R. Ballantyne\altaffilmark{1} and C. Papovich\altaffilmark{2}}

\altaffiltext{1}{Department of Physics, The University of Arizona,
  1118 East 4th Street, Tucson, AZ 85721; drb@physics.arizona.edu}
\altaffiltext{2}{\spitzer\ Fellow, Steward Observatory, The University
  of Arizona, 933 N.\ Cherry Avenue, Tucson, AZ 85721; papovich@as.arizona.edu}

\begin{abstract}
We present the results of calculations of the cosmic AGN background
spectrum from 3~keV ($4\times 10^{-4}$\micron) to 1000\micron. These
computations make use of the measured X-ray luminosity function and
its evolution, as well as fits from synthesis models of the cosmic
X-ray background (CXB) to predict the AGN contribution to the cosmic
infrared background (CIRB) for different models of the location and
distribution of the absorbing material. By comparing our results to
observational constraints we conclude that the current deep \spitzer\
surveys can account for the entire AGN contribution to the CIRB at
24\micron, but these AGN are almost all Compton-thin. In fact, the
calculations show that Compton-thick AGN are a small fraction of the
CIRB for $\lambda < 100$\micron. For this reason, the most efficient
method of identifying the Compton-thick AGN population is through hard
X-ray imaging at energies $\gtrsim 40$~keV. Selection of
AGN based on their rest-frame near-IR colors will miss low luminosity
type~2 AGN due to contamination from the host galaxy. Finally, the AGN
that dominate the CXB must have star formation rates $<
100$~M$_{\odot}$~yr$^{-1}$, consistent with them having similar
properties as the sources which dominate the CIRB at $z \sim
1$. Including the estimated re-radiated emission from star formation,
AGN and their host galaxies may contribute $\sim 30$\% of the CIRB at
70\micron, dropping to $\sim 10$\% at 24\micron\ and $\sim 1$\% at
1--10\micron.
\end{abstract}

\keywords{diffuse radiation --- galaxies: active --- galaxies:
  evolution --- galaxies: formation --- infrared: galaxies ---  X-rays:
  galaxies}

\section{Introduction}
\label{sect:intro}
The extragalactic cosmic background radiation encodes within it the
history of the formation and growth of galaxies
and stars, and thus elucidating its origin is a vital step toward
understanding these processes. In recent years significant progress
has been made in the study of the X-ray and infrared backgrounds
(hereafter, CXB and CIRB, respectively).  The CXB is now understood to arise from
predominately absorbed, or type 2, active galactic nuclei (AGN) over a
large range of luminosity, redshift and absorbing column density
\citep{sw89}. The CIRB \citep{hd01}
is dominated by the thermal dust emission of star-forming galaxies
over a wide range of redshift \citep{lpd05}. Despite their disparate origins
and energy ranges, there is an intimate connection between the CXB and
CIRB since the X-rays absorbed by the obscured AGN which dominate the
CXB will be thermalized and re-radiated in the IR. Thus, a potentially
significant fraction of the CIRB could be made up by this reprocessed
accretion energy. If most of the accretion onto black holes over the
history of the Universe has been obscured \citep[e.g.,][]{fi99} then a precise
measurement of the AGN contribution to the CIRB is required to
understand the growth of the supermassive black holes. Similarly, as
the CIRB is dominated by star-forming galaxies with the longer
wavelength emission resulting from galaxies at high redshifts \citep{lpd05}, a
determination of the fraction of emission from AGN as a function of
wavelength is required to accurately measure the history of star and galaxy
formation over cosmic time. Given a procedure to connect the X-ray
absorption to the expected IR emission one can use the information in the CXB
to predict the expected AGN contribution to the CIRB.

\citet{silva04} made use of the latest X-ray luminosity function (LFs) by collecting
observed IR spectral energy distributions (SEDs) for AGN with a given
X-ray luminosity and column density. They were then able to predict
the AGN contribution to the CIRB by integrating over the X-ray LF and
observed local $N_{\mathrm{H}}$ distribution. The strength of this
procedure is that it is based on observations of real AGN, but because
of this, \citet{silva04} were limited to small numbers of objects for
each of the their luminosity and $N_{\mathrm{H}}$ bins and thus are
possibly subject to the significant object-to-object variability
observed in AGN in the IR \citep[e.g.,][]{weed05}. In addition, due to the lack of high
angular resolution observations in the mid-IR, \citet{silva04} were
forced to extrapolate the observed SEDs beyond 20\micron. An alternative
approach was presented by \citet{trei04,tre06} who used theoretical IR dust
emission models to compute the expected IR SED for a given
X-ray absorption column and luminosity. These models included a
detailed radiative transfer calculation and will be accurate in the
mid-IR, but do not include gas and therefore cannot be easily
connected to the X-ray absorption properties of AGN.

Here, we present the first prediction of the AGN contribution to the entire
extragalactic background light from 3~keV to 1000~\micron. This
AGN background spectrum is based on fits to the CXB and therefore
automatically gives the AGN fraction to the CIRB over its entire
wavelength range. The calculation procedure is described in the next
section, and we present the results and discuss their implications in
Sect.~\ref{sect:discuss}. The standard first-year \textit{WMAP}
$\Lambda$-dominated cosmology is assumed throughout this paper, i.e.,
$H_0=70$~km~s$^{-1}$~Mpc$^{-1}$, $\Omega_{\Lambda}=0.7$, and
$\Omega_{m}=0.3$ \citep{spe03}.

\section{Calculations}
\label{sect:calc}
The calculations of AGN SEDs were performed using the photoionization
code Cloudy v.\ 05.07.06 \citep{fer98} and are described in detail by
\citet{bal06b}. This technique has the advantage that it
self-consistently treats the atomic gas physics along with the
detailed dust radiation physics (including PAHs and emission from very
small grains). In addition, physical properties of the medium
attenuating the AGN, such as its distance from the central engine, can
be varied allowing for possible constraints to be placed on important
parameters. Although the IR radiative transfer used in these
calculations is much less sophisticated than those used by
\citet{tre06}, \citet{bal06b} showed that the SEDs, when averaged over
a $N_{\mathrm{H}}$ distribution, have very similar properties to the
ensemble of AGN found in the deep surveys of \chandra, \xmm\ and
\spitzer. Furthermore, to construct the AGN background spectrum the
individual $N_{\mathrm{H}}$-averaged SEDs are integrated over all
luminosities and redshifts. This procedure further reduces the impact of
the simplified radiative transfer treatment.

To calculate the observed SED of an AGN with X-ray luminosity $L_X$
attenuated by a column density $N_{\mathrm{H}}$, we first define a model AGN continuum
extending from 100~keV to $>1000$~\micron\ using the
`agn' command in Cloudy. This spectrum has a X-ray photon index of
$\Gamma=1.9$ that falls off as $\nu^{-3}$ at energies $\geq
100$~\kev, a big blue bump temperature of $1.4\times 10^5$~K (i.e., a
$10^7$~M$_{\odot}$ black hole accreting at 10\% of its Eddington
rate), and $\alpha_{\mathrm{ox}}=-1.4$. Although this SED is a good
phenomenological description of an AGN spectrum, it is missing the
Compton reflection component in the hard X-ray band
\citep[e.g.,][]{np94} and has a lower cutoff energy than what is
typically observed in Seyfert galaxies \citep{m01}. The reflection
component and a high energy cutoff are both necessary to produce good
fits to the peak of the CXB at $\sim 30$~\kev\ \citep{com95,gch06} and
so the background spectrum computed with this SED will underpredict
the peak intensity of the CXB. However, as described below, the AGN
background spectrum is calculated using the population parameters
(i.e., evolution of type 2/type 1 ratio, hard X-ray luminosity function, $N_{\mathrm{H}}$
distribution) that were all used in fits to the peak of
the CXB \citep{bal06a}. In addition, the omission of the reflection
component will not affect the dust heating in the obscuring
medium. Therefore, the low energy background, including the CIRB, will
still be self-consistently predicted using the properties of a known
fit to the CXB.

This AGN spectrum illuminates a constant density ($n_{\mathrm{H}} =
10^4$~cm$^{-3}$) cloud of gas and dust with an inner radius either
$r_{\mathrm{in}}=1$ or $10$~pc from the continuum source.  The radial
extent of the irradiated cloud is determined by requiring the X-ray
spectrum to be absorbed by a column $N_{\mathrm{H}}$. The transmitted
incident spectrum, outwardly directed diffuse emission and reflected
radiation are combined under the assumption of the unified model to
produce a 'unified' SED for each $L_X$ and $N_{\mathrm{H}}$
\citep{bal06b}. Finally, at every $L_X$ and $z$, the unified SEDs
undergo a weighted average over $N_{\mathrm{H}}$ to produce the
$N_{\mathrm{H}}$-averaged SEDs that are then used to calculate the AGN
background spectrum. The weights are given by the distribution of
column densities predicted by the AGN type 2/type 1 ratio at the given
$L_{X}$ and $z$.

We employ a model of an evolving AGN type 2/type 1 ratio, such that
the fraction of type 2s with $41.5 \leq \log L_X \leq 48$ and $0 \leq
z \leq 1$ is determined by $f_2 = K (1 +z)^{0.3}(\log L_X)^{-4.8}$,
where $K$ is determined by requiring the low-$L_X$, $z=0$ type 2/type
1 ratio to be 4 \citep{mr95}. Between $z=1$ and $5$ the value of $f_2$
remains at its $z=1$ value for the appropriate $L_X$. \citet{bal06a}
found that this evolution can self-consistently account for the
spectrum of the CXB, the observed X-ray number counts and the values
of $f_2$ observed by \citet{bar05}. This redshift evolution of $f_2$
agrees with the recent measurements of \citet{tu06}, as well as the
luminosity evolution measured by \citet{gch06}. However, because the
AGN background spectrum is calculated by integrating the individual
$N_{\mathrm{H}}$-averaged spectra over all $L_X$ and $z$, the specific
evolution model of $f_2$ makes little impact to the final
result. Indeed, there is negligible difference between the background
spectrum calculated with the above evolution of $f_2$ and one where
$f_2$ remains constant with redshift.

AGN SEDs were computed with Cloudy for the following column densities: $\log
N_{\mathrm{H}} = 20, 20.5,\ldots,24.0,24.5$. Any object observed
through a column $\log N_{\mathrm{H}} \geq 22$ is considered a type~2
AGN. The distribution of column densities thus depends on luminosity
and redshift. We consider two different methods of determining the
$N_{\mathrm{H}}$ distribution. In the first scenario, any type~1 or 2
AGN has an equal probability $p$ of being assigned any column within
their classification: 
\begin{equation}
N_{\mathrm{H}} = \left\{ \begin{array}{lll}
                         20.0,\ldots,21.5 & & p=(1-f_2)/4.0\\
                         22.0,\ldots,24.5 & & p=f_2/6.0
                         \end{array}
                 \right .
\label{eq:simple}
\end{equation}
In the second method, the type~2 objects are distributed as observed
in local Seyfert~2 galaxies with a higher fraction of Compton-thick
sources \citep{rms99}:
\begin{equation}
N_{\mathrm{H}} = \left\{ \begin{array}{lll}
                         20.0,\ldots,21.5 & & p=(1-f_2)/4.0\\
                         22.0,\ldots,23.5 & & p=f_2/8.0\\
                         24.0,24.5 & & p=f_2/4.0
                         \end{array}
                 \right .
\label{eq:risaliti}
\end{equation}

The AGN background spectrum in $\nu I_{\nu}$ is then given by 
%
%
%
\begin{equation}
\begin{array}{ll}
\nu I_{\nu}(\lambda) = {c \over H_0} & \\
 & \\
\int_{z_{\mathrm{min}}}^{z_{\mathrm{max}}} \int_{\log
  L_{X}^{\mathrm{min}}}^{\log L_{X}^{\mathrm{max}}} {d\Phi(L_X,z) \over d\log L_
X}
  {S_{\lambda}(L_X,z) d_l^2 \over (1+z)^2 (\Omega_m (1+z)^3 +
  \Omega_{\Lambda})^{1/2}} d\log L_X dz, & \\
\end{array}
\label{eq:cxrb}
\end{equation}
where $d\Phi(L_X,z)/d\log L_X$ is the \citet{ueda03} X-ray luminosity function for AGN
(in Mpc$^{-3}$), $S_{\lambda}(L_X,z)$ is the observed-frame AGN
$N_{\mathrm{H}}$-weighted SED (in $\nu f_{\nu}$ units) calculated with
luminosity $L_X$ at redshift $z$, and $d_l$ is the luminosity distance
to redshift $z$ (where $z_{\mathrm{min}}=0$ and $z_{\mathrm{max}}=5$). AGN background spectra were calculated for
absorber radii of $r_{\mathrm{in}}=1$ and $10$~pc and with both the simple and
\citet{rms99} $N_{\mathrm{H}}$ distribution.

\section{Results and Discussion}
\label{sect:discuss}
\subsection{The AGN Background Spectrum}
\label{sub:bgrdspect}
\begin{figure*}
\begin{center}
\includegraphics[angle=-90,width=0.825\textwidth]{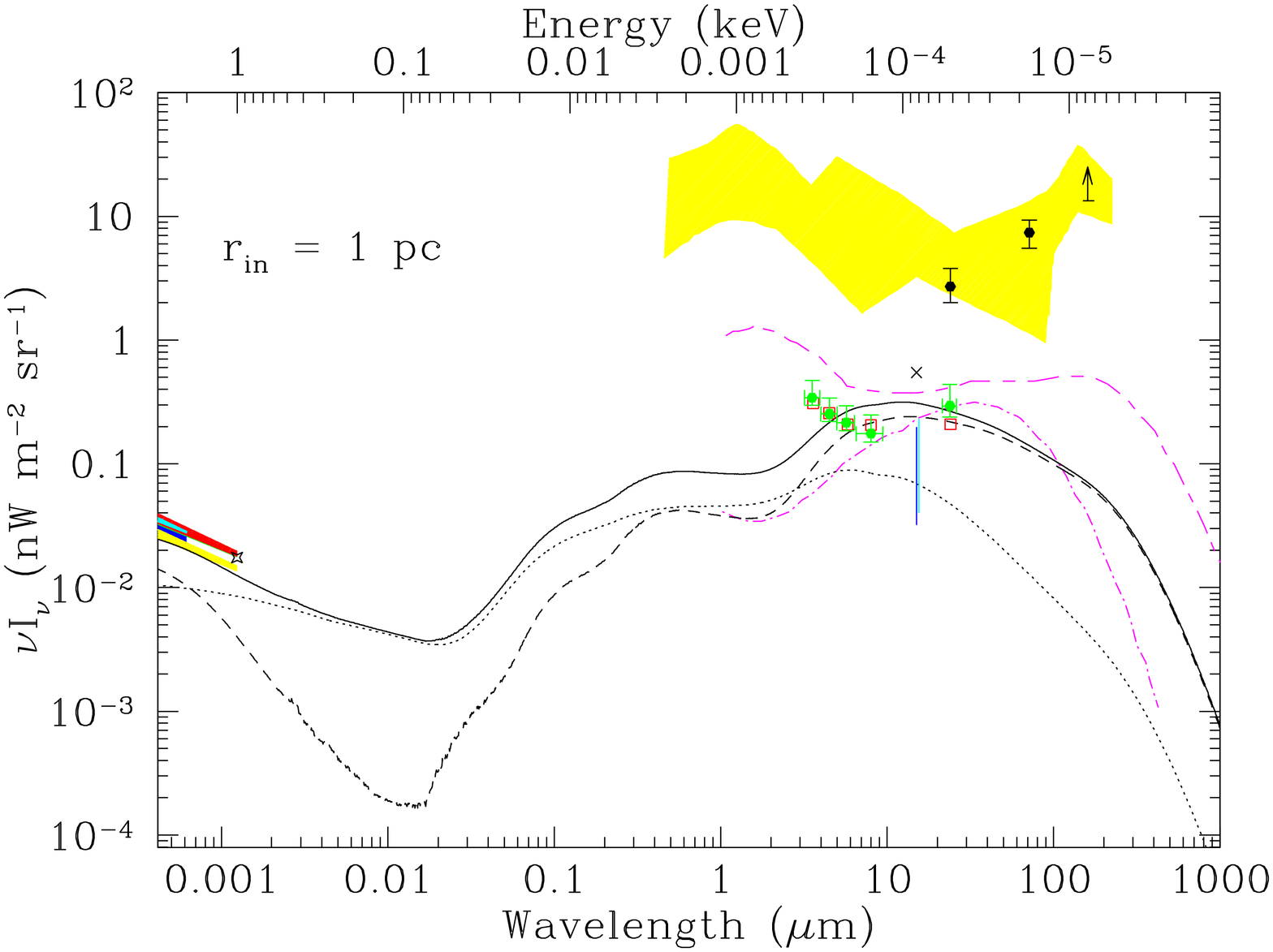}
\includegraphics[angle=-90,width=0.825\textwidth]{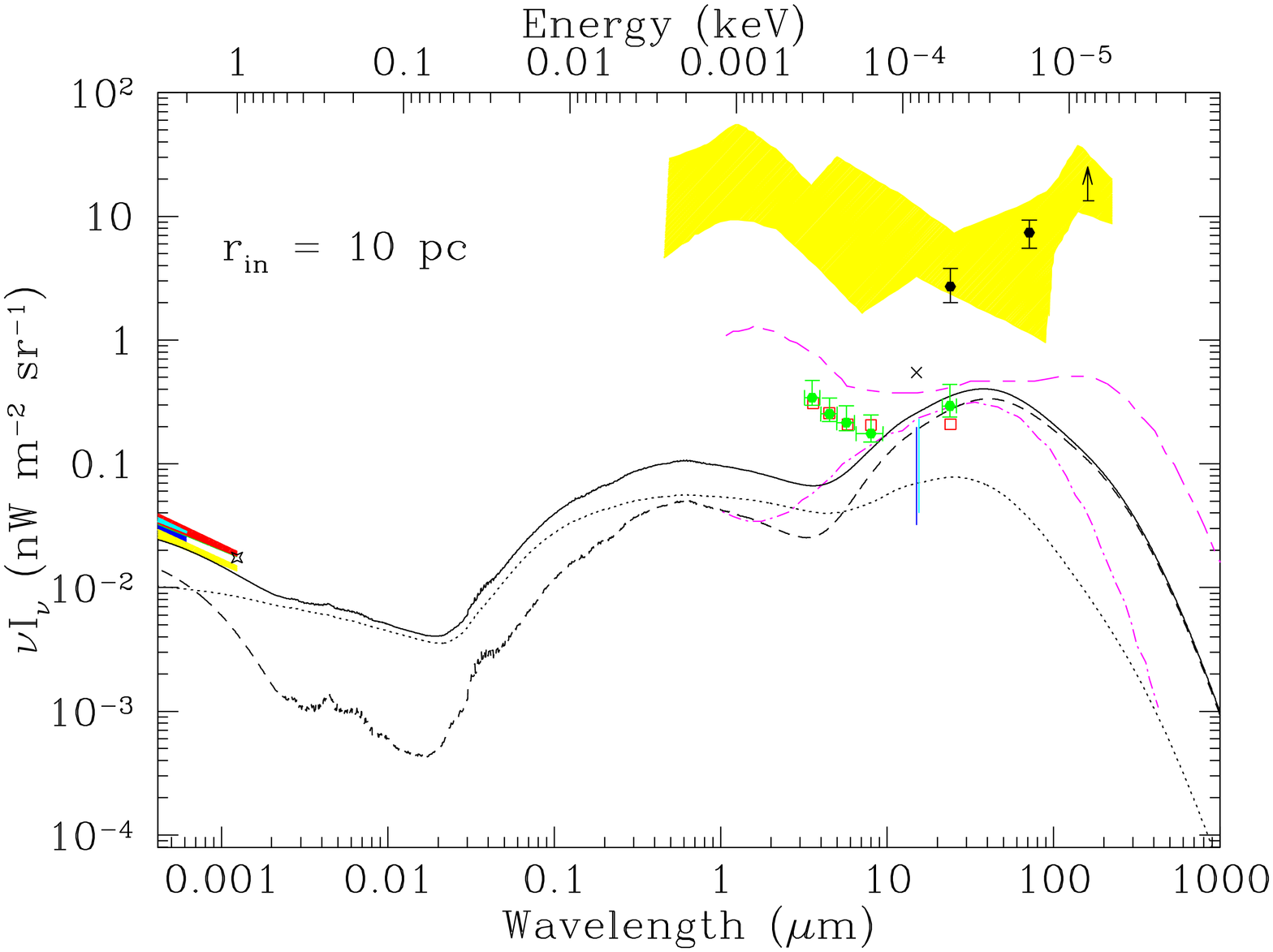}
\caption{(Top) The heavy solid line plots the predicted total AGN
  background spectrum from hard X-rays to the far-IR under
  the assumptions of $r_{\mathrm{in}}= 1$~pc and the simple
  $N_{\mathrm{H}}$ distribution. The dotted and
  dashed spectra are the contributions from type~1 ($\log
  N_{\mathrm{H}} < 22$) and type~2 AGN ($\log
  N_{\mathrm{H}} \geq 22$),
  respectively. In the X-ray band the
  different colored regions denote the CXB intensities measured from
  multiple missions (blue, \textit{ASCA} GIS, \citealt{kush02}; red, \textit{BeppoSAX},
  \citealt{vec99}; yellow, \textit{ASCA} SIS, \citealt{gen95}; cyan,
  \textit{XMM-Newton}, \citealt{dm04}). The open star plots the recent
  measurement of the intensity of the CXB at 1~\kev\ by \citet{hm06}. In the IR, the dot-dashed and dashed magenta lines plot the
  predicted contribution to the CIRB from AGN and AGN + host galaxies,
  respectively, calculated by \citet{silva04}. The green points and red squares
  are \spitzer\ measurements of the AGN contribution to the CIRB
  presented by \citet{tre06} and \citet{barm06}, respectively. The
  blue and cyan lines at 15\micron\ denote the range covered by type~1
  and type~2 AGN as determined by deep \textit{ISO} surveys
  \citep{mat06}. The black cross is an estimate from \citet{fad02} of
  the upper-limit to the AGN contribution to the CIRB at
  15\micron. The black circles are \spitzer\ measurements of the total
  CIRB at 24\micron\ \citep{pap04} and 70\micron\ \citep{fray06}. The
  stacking analysis of \citet{dol06} provides a lower-limit at
  160\micron. Finally, the yellow area shows the allowed range of the
  total CIRB as compiled by \citet{hd01}. (Bottom) As in the top
  panel, but now plotting a model where $r_{\mathrm{in}}= 10$~pc.}
\label{fig:cirb}
\end{center}
\end{figure*}

Figure~\ref{fig:cirb} plots the predicted AGN background spectrum from
3~keV to 1000\micron\ along with both recent observational
constraints on the CXB and CIRB and the models of \citet{silva04}.
Between 1 and 3~\kev\ the background model has the correct shape, but
falls at the lower edge of the range of measured CXB intensities. This is also found in the most recent synthesis models
\citep{gch06} and is likely a result of contributions from non-AGN sources
such as clusters and star-forming galaxies.

The dotted and dashed lines denote the contributions to the total AGN
background spectrum from type~1 ($\log N_{\mathrm{H}} < 22$) and
type~2 AGN ($\log N_{\mathrm{H}} \geq 22$), respectively. In
agreement with previous work \citep{silva04,trei04,bal06a}, the type~2
AGN dominate the AGN
background light at hard X-rays and in the mid- and far-IR. Type~1 AGN
only dominate the AGN background where the type~2 sources suffer
the greatest extinction. The predicted type~1 background spectrum passes through
the range of values measured by \textit{ISO} from type~1 AGN at
15\micron\ \citep{mat06}. This result implies that the deep \textit{ISO} surveys have
observed the vast majority of the type~1 AGN contribution to the CIRB
at 15\micron, although difficulties in identification resulted in
underestimating the fraction due to type~2 objects \citep{mat06}.

The two panels of Fig.~\ref{fig:cirb} show the predicted AGN
background spectra for $r_{\mathrm{in}}=1$~pc (top) and
$r_{\mathrm{in}}=10$~pc (top). It is interesting to compare the shapes
of the spectra in the IR band with those computed by \citet{silva04}
based on observations of nearby AGN. As our models do not include any
emission from the host galaxy, the appropriate comparison is with the
dot-dashed magenta line which plots the results of \citet{silva04} for
AGN with little to no host galaxy contamination. It is clear that the
$r_{\mathrm{in}}=10$~pc model provides a closer match to the
\citet{silva04} results (for $\lambda < 40$\micron) with the mid-IR peak occurring at nearly
identical wavelengths. The computed CIRB for this model predicts
significantly more emission beyond $\gtrsim 40$\micron\ than the
\citet{silva04} spectrum, but this is because these
authors extrapolated the observed SEDs to rest-frame wavelengths $> 20$\micron. In recent work, \citet{bal06b} found that,
based on comparisons with the observed \spitzer\ mid-IR number counts
and LFs, average AGN found in the X-ray and mid-IR deep surveys
were better described by a model where the obscuring medium around the
AGN is $\sim 10$~pc from the central engine as opposed to $\sim
1$~pc. Fig.~\ref{fig:cirb} indicates that the local AGN that
\citet{silva04} used to produce their CIRB prediction are also better
described by a model where the obscuring medium around the AGN is
$\sim 10$~pc from the central engine. These independent methods together
strongly imply that a $\sim 10$~pc scale obscuring medium is more
likely than smaller scales in AGN between $z=0$ and $1$.

\begin{figure*}
\begin{center}
\includegraphics[angle=-90,width=0.95\textwidth]{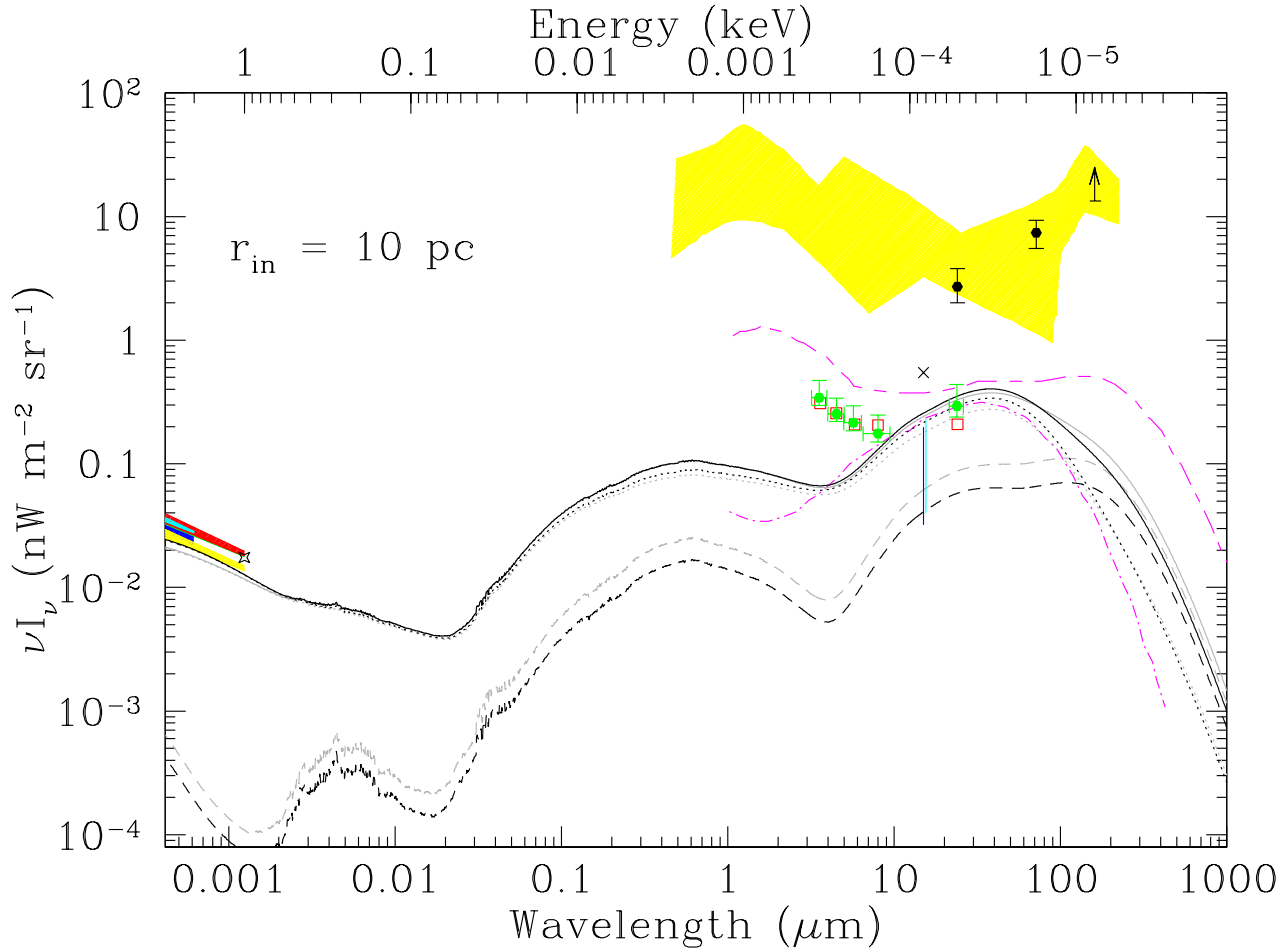}
\end{center}
\caption{As in the lower panel of Figure~\ref{fig:cirb}, but now dotted and
  dashed lines plot the contributions of Compton-thin ($\log
  N_{\mathrm{H}} < 24$) and
  Compton-thick AGN ($\log N_{\mathrm{H}} \geq 24$), respectively, to the total 
AGN background spectrum. The black lines show the results assuming the
  simple $N_{\mathrm{H}}$ distribution (eq.~\ref{eq:simple}). The gray
  lines plot the results when the \citet{rms99} $N_{\mathrm{H}}$
  distribution (eq.~\ref{eq:risaliti}) is assumed.}
\label{fig:cthick}
\end{figure*}

Focusing in on the $r_{\mathrm{in}}=10$~pc AGN background spectrum, we
see that the model is close to the observed \spitzer\ estimates for
the AGN contribution to the CIRB at 24\micron\
\citep{tre06,barm06}. Comparisons to data at shorter wavelengths are
complicated by the important host galaxy contamination that \spitzer\
will be be unable to avoid at faint IR fluxes (i.e., high-$z$
objects). This will result in an overestimation of the contribution to
the CIRB due solely from reprocessed light from the AGN. The host
galaxy contamination becomes increasingly important at shorter
wavelengths as a greater contribution from the Raleigh-Jeans tail of
the old stellar population within the galaxy enters into the
observations. This effect can be clearly seen at 3.6, 4.5, 5.7 and
8\micron\ in Fig.~\ref{fig:cirb}. This effect will undoubtedly cause
methods of AGN identification through near-IR properties \citep{ah06}
to miss type~2 AGN. That is, near-IR selection of AGN only works when
the re-radiated light from the AGN significantly outshines the host
galaxy in the rest-frame near-IR. Longer wavelengths, however, will be
immune to this problem so that a simple comparison can be made between
the \spitzer\ measurements at 24\micron\ and the model. Thus, we find
that the estimated contribution of AGN to the CIRB at 24\micron\ from
the GOODS survey (green dot; \citealt{tre06}) is consistent with the
background model and is unsurprisingly dominated by type~2 AGN. In
contrast, the measurement from the Extended Groth Strip (EGS;
\citealt{barm06}) lies below the background prediction by a factor of
1.7, and is therefore likely missing type~2 AGN. This difference is due to the
different survey depths. In the GOODS survey, deep \spitzer\
observations detected $> 90$\% of the sources found in the \chandra\ 2~Ms
deep fields. The shallower EGS survey
resulted in \spitzer\ detecting only $68$--$80$\% of the X-ray sources
and therefore underestimates the AGN contribution to the CIRB.

\subsection{The Importance of Compton-Thick AGN}
\label{sub:data}
An important problem in the study of both the CXB and CIRB is the
identification of Compton-thick AGN, objects with obscuring columns
$\log N_{\mathrm{H}} \geq 24$. Such heavily buried AGN are all but
invisible in the 2--10~\kev\ X-ray band \citep{matt99}, but a not insignificant
population of them are required in order to fit the peak of the CXB at
$30$--$40$~\kev\ \citep{gch06}. It has been hoped that mid-IR
observations will aid in the identification of Compton-thick AGN
\citep[e.g.,][]{don05}. Figure~\ref{fig:cthick} re-plots the $r_{\mathrm{in}}=10$~pc
AGN background spectrum from Fig.~\ref{fig:cirb}, but now shows the
contribution from Compton-thin ($\log N_{\mathrm{H}} < 24$) and
Compton-thick AGN. We see that Compton-thin AGN dominate the
production of the AGN background at all wavelengths $\lesssim
100$\micron. In our models, the heavily obscured Compton-thick objects
predominantly produce emission from cooler dust as the warm interior
is shielded from view \citep{bal06b}. Therefore, the predictions using
the \citet{rms99} $N_{\mathrm{H}}$ distribution, in which half of the
type 2 AGN are Compton thick (gray lines in Fig.~\ref{fig:cthick}),
find a slightly weaker CXB at $1$--$3$~\kev\ as that emission is
converted into enhanced far-IR radiation. This X-ray deficit would
have to be filled in with additional emission from other sources, or
through a revision in the hard X-ray LF.

At 24\micron\ the spectral intensity from Compton-thin AGN is $\sim
5\times$ greater than the Compton-thick sources. Assuming the
\citet{rms99} $N_{\mathrm{H}}$ distribution, this factor
reduces to about 3, although \citet{bal06a} found that the
\citet{rms99} distribution cannot hold over all $L_X$ and $z$ as it
underpredicts the observed 2--8~keV number counts. Thus the true
factor is likely between these two values. The estimated contribution
of AGN to the CIRB at 24\micron\ from the EGS survey lies
below the model prediction for the Compton-thin AGN contribution in
either $N_{\mathrm{H}}$ distribution, implying that this survey is
missing Compton-thin objects. In contrast, the measurement derived
from the GOODS survey is consistent with our model for the
Compton-thin objects at 24\micron. Thus, we conclude that in a deep
survey such as GOODS, the Compton-thin AGN are basically completely
identified and these AGN constitute the vast majority of the AGN
background at 24\micron. Interestingly, our models show that
Compton-thick AGN comprise only a small fraction of the AGN background
at this wavelength (Fig.~\ref{fig:cthick}). Recent analysis of the CXB
has shown that the deep X-ray surveys are missing $\sim 50$\% of the
high-$L_X$ obscured AGN \citep{wor05,bal06a}, many of which are
expected to be Compton-thick \citep{gch06}. Fig.~\ref{fig:cthick}
suggests that using the mid-IR to identify such highly obscured
objects will be extremely difficult as they constitute only a tiny
fraction of the AGN background at these wavelengths. Compton-thick AGN
do dominate the total AGN emission in the far-IR ($\gtrsim
100$\micron), but the CIRB reaches its peak in this region due to
emission from star formation over a wide range of $z$ \citep{lpd05},
and the AGN contribution is therefore a very small fraction of the
total radiation emitted by star-forming galaxies. Thus, we conclude
that, irrespective of the exact fraction of Compton thick AGN, they
contribute only a small fraction to the total AGN extragalactic
background light except for in the far-IR and the very hard X-ray band
\citep{gch06}. Therefore, the best strategy to find Compton-thick AGN
is through X-ray imaging at energies $\gtrsim 40$~keV.

\subsection{Accounting for Star-Formation}
\label{sub:stars} 
\begin{figure*}
\begin{center}
\includegraphics[angle=-90,width=0.95\textwidth]{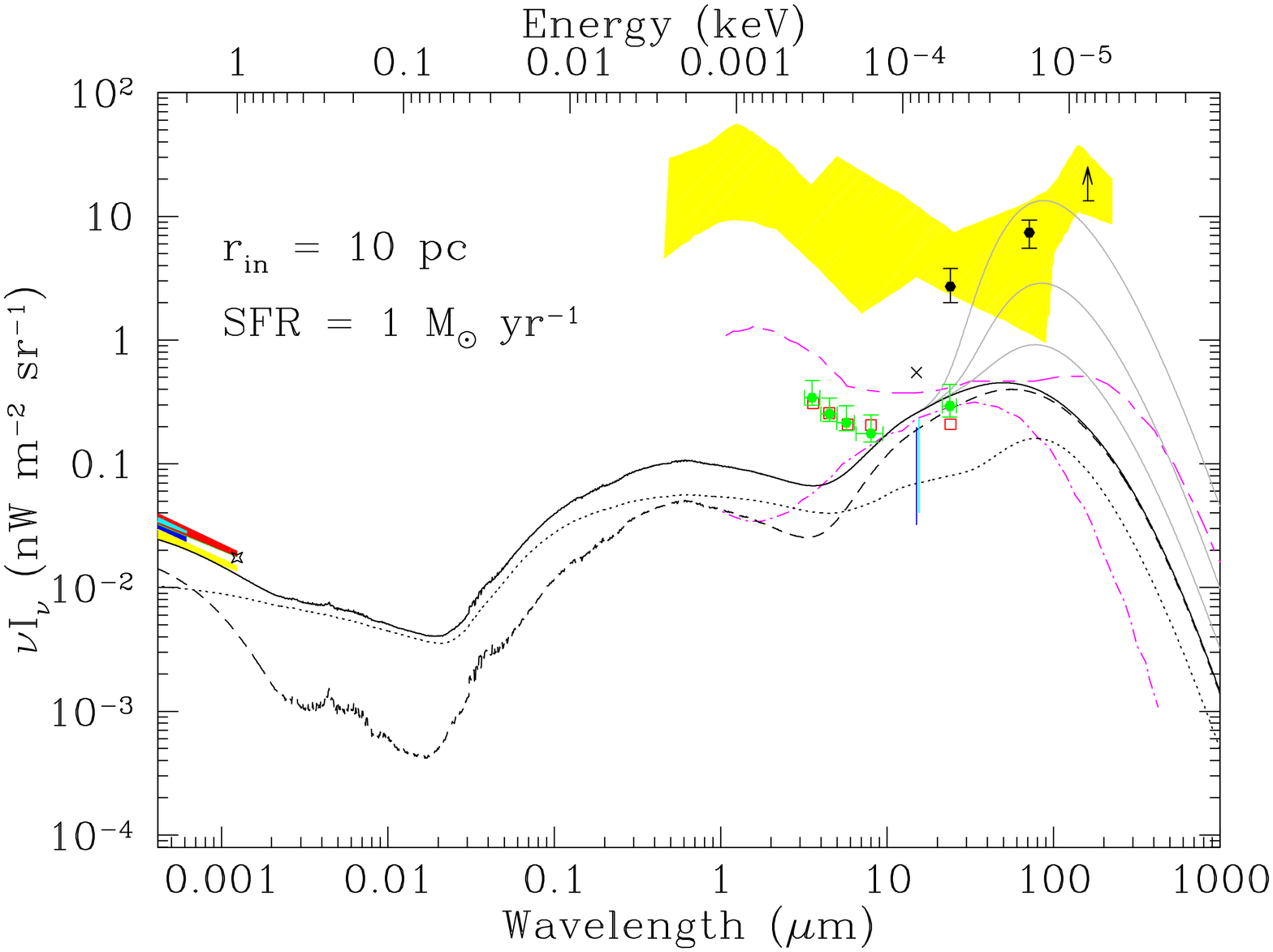}
\end{center}
\caption{As in the bottom panel of Fig.~\ref{fig:cirb}, but
  now including an estimate for the IR emission between 3 and
  1500\micron\ (rest-frame) from reprocessed light due to an
  instantaneous star-formation rate of of 1 (black solid line), 5, 20 and
  100~M$_{\odot}$~yr$^{-1}$ (gray solid lines). These calculations
  used the \citet{yc02} estimate for the IR continuum in star-forming
  galaxies, i.e., a constant dust temperature of $T_{\mathrm{d}} =
  58$~K. These results also assumed the simple $N_{\mathrm{H}}$
  distribution.}
\label{fig:cirb-sf}
\end{figure*}

Along with emission from old stars in the AGN host galaxy, reprocessed
light from star-forming regions emitted in the mid- and far-IR can be
an important contributor to the IR spectra of AGN. \citet{yc02}
derived an empirical formula relating the instantaneous star-formation
rate (SFR) and the observed mid-to-far IR continuum in star-forming
galaxies. Fig.~\ref{fig:cirb-sf} shows the results of including this
simple constant temperature ($T_{\mathrm{d}} = 58$~K; \citealt{yc02})
estimate for the effects of star-formation on the AGN background
spectrum assuming an instantaneous SFR of $1$ (solid black line), $5,
20$ and $100$~M$_{\odot}$~yr$^{-1}$ (sold gray lines). This
star-formation spectrum was added to each AGN SED between 3 and
1500\micron\ (rest-frame) prior to performing the background integral
(eq.~\ref{eq:cxrb}). The galaxies fit by \citet{yc02} all
have SFRs$ >20$~M$_{\odot}$~yr$^{-1}$, so the dust temperatures for
the $1$ and $5$~M$_{\odot}$~yr$^{-1}$ models will likely be
overestimates. However, this would only move the peak to the right on
the plot and not greatly affect the amplitude.

The dashed magenta line in Fig.~\ref{fig:cirb-sf} plots the
\citet{silva04} model of the AGN contribution to the CIRB including
emission from host galaxies. The maximum SFR allowed by the
\citet{silva04} result is only $1$~M$_{\odot}$~yr$^{-1}$. The
\citet{silva04} curve peaks at longer wavelengths than our model, but
this is just a result of the high dust temperature (for this SFR) used
in our estimate. The match in amplitude is still robust. Thus, if this
is generally correct, the typical SFR in an average AGN over all
luminosities and $z$ is $\sim 1$~M$_{\odot}$~yr$^{-1}$. However, only
low-$z$ AGN were used by \citet{silva04} to construct their
SEDs. Indeed, the average luminosity of their AGN host galaxy sample
is $L_{\mathrm{IR}} = 10^{10}$~L$_{\odot}$, which is at the threshold of
starbursts, and if entirely attributed to star formation corresponds
to a SFR of $\sim 1$~M$_{\odot}$~yr$^{-1}$. The CXB, in contrast, is mostly produced at $z \sim 1$, where
the cosmic SFR density is $\sim 10\times$ greater \citep{hopk04}. The
seeming increase in the AGN type~2 fraction with $z$ also indicates
that the SFR may be higher in these objects
\citep{bar05,bal06a,tu06}. At such redshifts, the CIRB is dominated by
the luminous IR galaxies with typical SFRs $\gtrsim
20$~M$_{\odot}$~yr$^{-1}$ \citep{lpd05}.

\begin{figure*}
\includegraphics[angle=-90,width=0.95\textwidth]{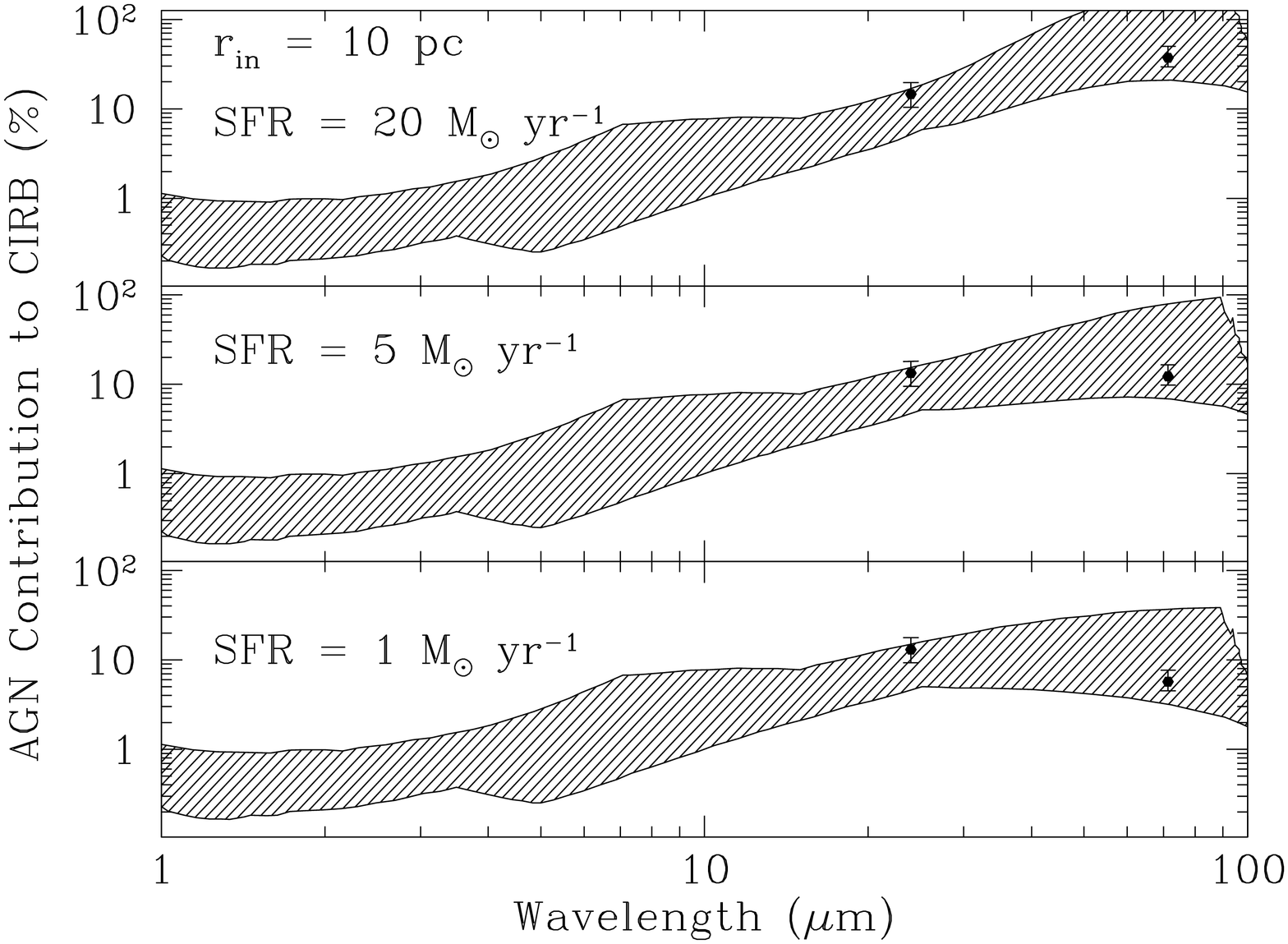}
\caption{The predicted contribution of AGN + reprocessed
  star-formation light to the CIRB using the $r_{\mathrm{in}}= 10$~pc,
  SFR$=20$~M$_{\odot}$~yr$^{-1}$, $5$~M$_{\odot}$~yr$^{-1}$ and
  $1$~M$_{\odot}$~yr$^{-1}$ models from Fig.~\ref{fig:cirb-sf}. The
  hatched region shows the range of possible percentages calculated by
  dividing the models by the yellow region from Fig.~\ref{fig:cirb-sf}
  \citep{hd01}. The data points at 24 and 70\micron\ were calculated
  by dividing the models by the \spitzer\ measurements of
  \citet{pap04} and \citet{fray06}, respectively.}
\label{fig:percent}
\end{figure*}

The percentage of the CIRB produced by AGN and star-formation in their
host galaxies will depend both on the SFR within the host galaxy and
the wavelength of interest. Fig.~\ref{fig:percent} plots the
percentage of the CIRB due to AGN + reprocessed star-formation light
between 1 and 100\micron. The hatched region shows the allowed
percentages when the models of Fig.~\ref{fig:cirb-sf} are divided by
the yellow region indicating the compilation of measurements by
\citet{hd01}. Similarly, the data points at 24 and 70\micron\ were
calculated by dividing the models by the \spitzer\ measurements of \citet{pap04} and
  \citet{fray06}. If AGN host galaxies do have significant SFRs on
  average, then $\sim 30$\% of the CIRB at
70\micron\ could be due to AGN + host galaxy emission (Fig.~\ref{fig:percent}). At shorter
wavelengths the percent of the CIRB produced by AGN depends only
slightly on the host galaxy SFR, and is $\sim 10$\% at 24\micron,
dropping to $\sim 1$\% between 1 and 10\micron.

Finally, Fig.~\ref{fig:cirb-sf} shows that AGN which produce the CXB
cannot, on average, have SFRs $\gtrsim
100$~M$_{\odot}$~yr$^{-1}$. Such very high SFRs are the hallmark of
the ultra-luminous IR galaxy population which are observed
to be associated with major merger events \citep{san88,san88b} and whose evolution 
in space density is very similar to quasars \citep{ks98,bar00}; that
is, they peak in density at $z \sim 2$--$3$. This result
may imply that the moderate-luminosity AGN at $z \sim 1$
that generate the CXB have different evolutionary paths than the
high-luminosity quasars and be fueled by different mechanisms
(minor mergers, interactions, etc.). Rather than signifying the rapid birth
of a large elliptical galaxy, these intermediate AGN are the signs of a
more leisurely mode of galaxy formation. Alternatively, these AGN
could be the fading remnants of the higher redshift quasars. However,
the large population of LIRGs at $z \sim 1$ indicates that there is
significant galaxy growth at this epoch which involves significant gas
being condensed in the nuclei of galaxies. The correlation between
black hole and bulge mass seen in spheroidal galaxies \citep{tre02} shows that
the star formation and black hole growth are connected during massive galaxy
formation at high $z$ \citep{dsh05}. Therefore, it is highly plausible that they may
be connected with reduced intensity at $z \sim 1$. Future measurements of the AGN
contribution to the CIRB should be able to constrain the SFR in AGN at
the peak of obscured AGN activity. Clearly, a detailed
study of the CXB and CIRB probes the intertwined processes of black
hole growth, galaxy evolution and star formation.

\acknowledgments

The authors thank the anonymous referee for very helpful comments, and
R.\ Gilli for kindly sending the CXB data.  DRB is supported by the
University of Arizona Theoretical Astrophysics Program Prize
Postdoctoral Fellowship. Support for this work was provided by NASA
through the \spitzer\ Space Telescope Fellowship Program, through a
contract issued by the JPL, Caltech under a contract with NASA.

{}

\end{document}